\documentclass[aps,showpacs,prep,graphics,twocolumn]{revtex4}

\usepackage{amssymb}
\usepackage[dvips]{graphicx}
\usepackage{amsmath}
\usepackage{bm}
\usepackage{epsfig}
\usepackage{graphicx}
\usepackage{caption}
\usepackage{wrapfig}
\usepackage{color}

\begin{document}

\title{ Non-perturbative gauge invariant scalar fluctuations  of the metric in Higgs inflation from complex geometrical scalar-tensor theory of gravity.}

\author{ $^{1}$ Jos\'e Edgar Madriz Aguilar\thanks{E-mail address: madriz@mdp.edu.ar}, $^{2}$ A. Bernal, $^{1}$ M. Montes, $^{3}$ J. Zamarripa and $^{3}$ E. Aceves} 
\affiliation{$^{1}$ Departamento de Matem\'aticas, Centro Universitario de Ciencias Exactas e ingenier\'{i}as (CUCEI),
Universidad de Guadalajara (UdG), Av. Revoluci\'on 1500 S.R. 44430, Guadalajara, Jalisco, M\'exico.\\
\\
$^{2}$ Departamento de F\'isica,\\
Centro Universitario de Ciencias Exactas e Ingenier\'ias (CUCEI),\\
Universidad de Guadalajara.\\
Av. Revoluci\'on 1500 S. R. 44430,\\
Guadalajara, Jalisco, M\'exico.
and\\
\\
$^{3}$ Centro Universitario de los Valles, Carretera Guadalajara-Ameca Km 45.5, C. P. 46600, Ameca, Jalisco, M\'exico.\\ \\
E-mail: jose.madriz@academicos.udg.mx, alfonso.bernal@alumnos.udg.mx, mariana.montes@academicos.udg.mx, jose.zamarripa@academicos.udg.mx,
elizabeth.aceves5374@alumnos.udg.mx }

\begin{abstract}

In this letter we investigate gauge invariant scalar fluctuations of the metric in a non-perturbative formalism for a Higgs inflationary model recently introduced in the framework of a geometrical scalar-tensor theory of gravity. In this scenario
the Higgs inflaton field has its origin in the Weyl scalar field of the background geometry. We found a nearly scale invariance of the power spectrum for linear scalar fluctuations of the metric. For certain parameters of the model we obtain values for the scalar spectral index $n_s$ and the scalar to tensor ratio $r$ that fit well with the Planck 2018 results. Besides we show that in this model the trans-planckian problem can be avoided.

\end{abstract}

\maketitle

\vskip .5cm
 Weyl-Integrable geometry, geometrical scalar-tensor gravity, Higgs cosmological inflation, gauge invariant fluctuations of the metric.

\section{Introduction}

The theory of cosmological perturbations is a very important part in the physical description of the inflatio\-nary epoch. It describes the formation and evolution of the seeds of cosmological structure. Some inflationary scenarios are compatible with measurements of cosmic microwave background (CMB) and Planck 2018 Results \cite{Infla1, Planck2018}. In the inflationary models the inflaton field generates the enough vacuum energy to solve the old problems of big bang cosmology. However, the unique scalar particle we have experimental evidence of its existence is the Higgs boson, which has been observed in 2012 with a mass of 125 GeV \cite{HiggsD1,HiggsD2}. This discovery has led several cosmologists to propose that the Higgs scalar field might be the same as the inflaton field \cite{HI1}. The main problem in regarding this idea is related with the so called ``hierarchy problem". This consists in the obervational fact that the Higgs boson mass seems to be sensitive to quantum corrections and the bare Higgs mass then need to be fine-tuned to achieve a physical Higgs boson mass many orders of magnitude smaller than the Planck scale \cite{Infla2}. It basically means that there is a big gap of energy between the electroweak and the Plack scale, and thus the Higgs field in this conditions results to be too small to generate the enough energy to inflate the primordial universe. In particular, in order to have the enough inflation to solve the big bang problems, the inflaton is estimated to have a mass $\sim 10^{13}$ {\it GeV} \cite{HI1A,HI1B}. Among many others, we can find in the literature models with non-minimally coupled inflaton Higgs field trying to alleviate this issue \cite{HI2,HI3,HI4,HI5}. Some other attempts include models in the Palatini approach \cite{Agr1,Agr2,Agr3,Agr4,Agr5,FR,SUR1,SUR2} and models in non-riemannian geometries \cite{Infla1,NRI}.\\

On the other hand, a way to introduce a scalar field in a gravitational framework in a geometrical manner is shown in the recently introduced geometrical scalar-tensor theories of gravity \cite{c10,c13}. These theories arose as an attempt to alleviate the Jordan to Einstein frame controversy. In this new approach the scalar-tensor theory is formulated in a non-riemannian geometry which is obtained via the Palatini variational principle.
In this framework the inflaton scalar field is introduced as a part of the affine structure of the geometry and thus it results to be related to Weyl scalar field \cite{c10,c13, NRI}. Some other topics like  $(2+1) $ gravity models, inflation and cosmic magnetic fields, quintessence and some cosmological models  have been studied in this approach  \cite{c14,H16,H17,H18}.\\

In this letter we use a non-perturbative a\-pproach to study linear scalar fluctuations of the metric in a recently introduced Higgs inflationary model \cite{NRI} developed in the context of a geometrical scalar-tensor theory of gravity. Thus, in section I we give a brief introduction. Section II is devoted to the basic formalism of geometrical scalar-tensor theories. Section III is left for the scalar fluctuations of the metric of arbitrary amplitude. In section IV we study linear fluctuations of the metric and we obtain their power spectrum. Finally, in section V we give some conclusions.

\section {Basic formalism}

We start by considering a complex scalar-tensor theory of gravity in vacuum whose action reads
 \begin{small}
 \begin{equation}\label{f1}
 {\cal S}=\frac{1}{16\pi}\int d^4x \sqrt{-g}\left[\tilde{\Phi}\tilde{\Phi}^\dagger {\cal R}+\frac{\tilde{W}(\tilde{\Phi}\tilde{\Phi} ^\dagger)}{\tilde{\Phi}\tilde{\Phi} ^\dagger}g^{\mu\nu}\tilde{\Phi}_{,\mu}\tilde{\Phi}_{,\nu}^{\dagger} -\tilde{U}(\tilde{\Phi}\tilde{\Phi} ^\dagger)  \right] 
 \end{equation}
\end{small}
where ${\cal R}$ denotes the Ricci scalar, $ \tilde{W}(\tilde{\Phi}\tilde{\Phi} ^\dagger)$ is a well-behaved differentiable function of $\tilde{\Phi}\tilde{\Phi} ^\dagger$, the dagger $\dagger$ denotes transposed complex conjugate  and $\tilde{U}(\tilde{\Phi}\tilde{\Phi} ^\dagger)$ is a scalar potential. By means of the transformation $\tilde{\Phi}=\frac{1}{\sqrt{G}}e^{-\varphi}$ the action \eqref{f1} can be recasted in the more convenient form
\begin{eqnarray}
{\cal S}=\int d^{4}x\sqrt{-g}e^{-(\varphi+\varphi^\dagger)}\left[\frac{\mathcal{R}}{16\pi G} + \hat{\omega}(\varphi+\varphi^\dagger)g^{\mu\nu}\varphi_{,\mu}\varphi^\dagger_{,\nu}\right.\nonumber\\
\left.- \hat{V}(\varphi+\varphi^\dagger)\right],\nonumber\\
\label{f2}
\end{eqnarray}
where we have made the identifications  $\hat{\omega}(\varphi+\varphi^\dagger)=(1/16\pi)\tilde{W}(\varphi+\varphi^\dagger)e^{\varphi+\varphi^\dagger}$ and $ \hat{V}(\varphi+\varphi^\dagger)=(1/16\pi)\tilde{U}(\varphi+\varphi^\dagger)e^{\varphi+\varphi^\dagger}$. Now, to determine the background geometry corresponding to the action \eqref{f2} we use the Palatini va\-ria\-tio\-nal principle. Thus we arrive to the compatibility condition 
\begin{equation}\label{for2}
\nabla_{\mu} g_{\alpha\beta}=(\varphi + \varphi^{\dagger})_{,\mu}g_{\alpha\beta}.
\end{equation}
Hence, the background geometry is Weyl-integrable. It must be noted that \eqref{for2} is invariant under the symmetry group of transformations
 \begin{eqnarray}\label{f4}
  \bar{g}_{\mu\nu}&=& e^{f+f^\dagger}g_{\mu\nu},\\
  \label{f5}
    \bar{\varphi}&=& \varphi +f,\\
    \label{f55}
     \bar{\varphi}^\dagger&=& \varphi^{\dagger} +f^\dagger,
  \end{eqnarray}
where $f=f(x^{\alpha})$ is a well defined complex function of the space-time coordinates. Unfortunately, as it was shown in \cite{NRI,H16,H17} the action \eqref{f2} does not remain invariant under the symmetry group of the geometry \eqref{f4}-\eqref{f55}. Thus it is proposed the action
\begin{eqnarray}
&&{\cal S}=\int d^4x\sqrt{-g}\,e^{-(\varphi+\varphi^\dagger)}\left[\frac{\mathcal{R}}{16\pi G}+\widehat{\omega}(\varphi+\varphi^\dagger) g^{\mu\nu}\varphi_{:\mu}\varphi^\dagger_{:\nu}\right.\nonumber\\
\label{f9}
&&\left.
-e^{-(\varphi+\varphi\dagger)}\widehat{V}(\varphi+\varphi^\dagger)\right],
\end{eqnarray}
where $\varphi_{:\mu}=\,^{(w)}\nabla_{\mu}\varphi+\gamma B_{\mu}\varphi$, is a gauge covariant derivative with  $B_{\mu}$ being a gauge vector field, $^{(w)}\nabla_{\mu}$ is the Weyl covariant derivative determined by (\ref{for2}) and  $\gamma$ is a pure imagi\-nary  coupling constant introduced to have the co\-rrect physical units. Notice that the action \eqref{f9} correspond to a non-conventional scalar-tensor theory of gravity. The invariance of \eqref{f9} under \eqref{f4}-\eqref{f55} is guaranteed once the vector field $B_{\mu}$, the function $\hat{\omega}$ and the scalar potential $\hat{V}(\varphi)$, obey the transformation rules
\begin{eqnarray}\label{f10a}
\bar{\varphi}\bar{B}_{\mu} &=& \varphi B_{\mu}-\gamma^{-1}f_{,\mu},\\
\label{f10aa}
\bar{\varphi}^\dagger\bar{B}_{\mu} &=& \varphi B_{\mu}+\gamma^{-1}f_{,\mu}^\dagger,\\
\bar{\hat{\omega}}(\bar{\varphi}+\bar{\varphi}^\dagger)&\equiv&\hat{\omega}(\bar{\varphi}+\bar{\varphi}^\dagger-f-f^\dagger)=\hat{\omega}(\varphi+\varphi^{\dagger}),\label{f10b}\\
\bar{V}(\varphi+\varphi^\dagger)&\equiv& V(\bar{\varphi}+\bar{\varphi}^\dagger-f-f^\dagger)=V(\varphi+\varphi^\dagger).\label{f10c}
\end{eqnarray}
Notice that \eqref{f10a} and (\ref{f10aa}) are transformation rules for the product $\varphi B_{\mu}$. Besides they have the same algebraic form of the elements of the Lie algebra associated to the group $U(1)$ employed in quantum electrodynamics. Thus, we may include a dynamics for $\varphi B_{\alpha}$ extending the action (\ref{f9}) by adding an electromagnetic type term in the form
\begin{small}
\begin{eqnarray}
&& {\cal S}=\int d^{4}x\sqrt{-g}\,e^{-(\varphi+\varphi^{\dagger})}\left[\frac{{\cal R}}{16\pi G}+\hat{\omega}(\varphi+\varphi^{\dagger})g^{\alpha\beta}\varphi_{:\alpha}\varphi_{:\beta}\right.\nonumber\\
&&\left.
-e^{-(\varphi+\varphi^{\dagger})}\hat{V}(\varphi+\varphi^{\dagger})-\frac{1}{4}e^{(\varphi+\varphi^\dagger)}H_{\alpha\beta}H^{\alpha\beta}\right],
\label{f14}
\end{eqnarray}
\end{small}
where  $H_{\alpha\beta}=(\varphi B_{\beta})_{,\alpha}-(\varphi B_{\alpha})_{,\beta}$ is the field strength asso\-ciated to the gauge boson field $B_{\mu}$.\\ 

The action (\ref{f14}) is an invariant action compatible with its background geo\-metry and originates a new kind of complex scalar-tensor theory of gravity. This action is written in terms of the metric $g_{\mu\nu}$ which according to Weyl transformations \eqref{f4} is not a Weyl-invariant. Moreover, the diferential line element transforms as
\begin{equation}\label{qnotas1}
d\bar{s}^2=e^{f+f^{\dagger}}ds^2.
\end{equation}
Thus, we introduce the Weyl-invariant metric
 \begin{equation}\label{qnotas2}
 h_{\mu\nu}\equiv e^{-(\varphi+\varphi^{\dagger})}g_{\mu\nu}.
 \end{equation}
In terms of this Weyl-invariant metric the action (\ref{f14}) acquires the form
\begin{eqnarray}
{\cal S}=\int d^{4}x\sqrt{-h}\left[\frac{\mathcal{R}}{16\pi G} +\hat{\omega} (\varphi+\varphi^\dagger) h^{\mu\nu}\mathbb{D}_\mu\varphi\mathbb{D}_\nu\varphi ^\dagger\right.\nonumber\\
\label{Rie2}
\left.-\widehat{V}(\varphi+\varphi^\dagger)-\frac{1}{4}H_{\mu\nu}H^{\mu\nu}\right],
\end{eqnarray}
where now the gauge covariant derivative becomes  $\mathbb{D}_{\mu}= \,^{(R)}\!\nabla_{\mu}+\gamma B_{\mu}$  and  the operator $^{(R)}\!\nabla_\mu$  denotes the Riemannian covariant derivative.\\

On the other hand, the non-metricity associated to the background geometry of the action \eqref{f1} is $N_{\mu\alpha\beta}=-[\ln(\tilde{\Phi}\tilde{\Phi}^{\dagger})]$. This non-metricity is quadratic in $\tilde{\Phi}$. However, when we implemented the transformation $\tilde{\Phi}=\frac{1}{\sqrt{G}}e^{-\varphi}$ the quadratic dependence in both the non-metricity and the action is lost, as shown in  \eqref{f2} and \eqref{for2}.  Thus, in order to restore the quadratic dependence in the scalar field, we introduce the field  transformations 
\begin{eqnarray}
\zeta &=&\sqrt{\xi}\,e^{- \varphi},\\
A_\mu&=&B_\mu \ln (\zeta/\sqrt{\xi}),
\end{eqnarray}
where $\xi$ is a constant introduced in order to the field $\zeta$ has the correct physical units. \\

Hence, the action (\ref{Rie2}) rewritten in terms of the fields $\zeta$ and $A_{\mu}$ becomes
\begin{eqnarray}
&& {\cal S}=\int d^{4}x\sqrt{-h}\left[\frac{\mathcal{R}}{16\pi G} +\frac{1}{2}\omega(\zeta\zeta ^\dagger) h^{\mu\nu} D _\mu \zeta( D _\nu \zeta) ^\dagger\right.\nonumber\\
\label{yq3}
&&
 \left. -V(\zeta\zeta ^\dagger)-\frac{1}{4}F_{\mu\nu} F^{\mu\nu}\right],
\end{eqnarray}
where $\mathcal{D}_\mu \zeta\equiv\zeta\mathbb{D}_\mu(\ln\frac{\zeta}{\sqrt{\xi}})=\,^{(R)}\nabla_{\mu}\zeta +\gamma A_{\mu}\zeta$ is an effective covariant derivative, $F_{\mu\nu}\equiv\partial_{\mu}A_{\nu}-\partial_{\nu}A_{\mu}=-H_{\mu\nu}$ is the Faraday tensor and  where we have made the identifications
 	\begin{eqnarray}
 	 \frac{\omega(\zeta \zeta^\dagger)}{2}&\equiv& \frac{\widehat{\omega}(\ln\frac{\zeta \zeta ^\dagger}{\xi})}{\zeta \zeta^\dagger},\label{5.46} \\
 	V(\zeta\zeta^\dagger)&\equiv &\widehat{V}\bigg(\ln\frac{\zeta\zeta^\dagger}{\xi}\bigg).\label{5.45}
 \end{eqnarray}
  The effective background geometry of (\ref{yq3}) is riemannian and it results invariant under the gauge transformations
 \begin{eqnarray}\label{5.48}
 \bar{\zeta}&=& \zeta e^{\gamma\theta(x)}\\
 \label{5.49}
 \bar{A}_\mu &=& A_\mu -\theta_{,\mu},
 \end{eqnarray}
 where $\theta(x)$ is a well-behaved function. Thus, the last term in \eqref{yq3} together with the transformations \eqref{5.48} and (\ref{5.49}), suggest that $A_\mu$ can play the role of an electromagnetic potential. However, it is important to note that the part of \eqref{yq3} that we relate with electromagnetism has its origin in the required Weyl invariance of the action \eqref{f9}.

 \section{Non-perturbative scalar fluctuations of the metric in a Higgs inflation model}
 
 In order to study gauge invariant scalar fluctuations of the metric, let us first give the basic formulation of a Higgs inflationary model derived from the formalism explained in the previous sections. In particular we will use the model proposed in \cite{NRI}. Thus we consider the Higgs potential in the Weyl frame in the form
 \begin{equation}\label{hpot1}
 \tilde{V}(\Phi\Phi^{\dagger})=\frac{\lambda}{4}\left(\Phi\Phi^{\dagger}-\sigma^2\right)^2,
 \end{equation}
 where accor\-ding to the best-fit experimental data  $\lambda=0.129$ and the vacuum expectation value for electroweak interaction $\sigma=246\,GeV$ \cite{expdata1,expdata2}. Thus, the Higgs potential in terms of the field $\zeta$ in the Riemann frame acquires the form
  \begin{equation}\label{5.50}
  V(\zeta\zeta^\dagger)=\frac{\lambda}{4}\left(\frac{\zeta\zeta^\dagger}{\xi}-\sigma^2\right)^2.
  \end{equation}
 The ground state $||\zeta\zeta^\dagger||=\sqrt{\xi}\sigma$ associated with \eqref{5.50} is inva\-riant under  \eqref{5.48}.  Howe\-ver, the breaking of the symmetry is achieved when we take $\zeta=\zeta^\dagger$ because in this particular case $||\overline\zeta^2||\neq||\zeta^2|$. Thus, excitations around the ground state read
 \begin{equation}\label{phiexp}
 \zeta(x^\mu)=\sqrt{\xi}\,\sigma+{\cal{Q}}(x^\mu),
 \end{equation}
 where ${\cal Q}(x)$ denotes the Higgs scalar field. 
 According to \eqref{phiexp} the kinetic term in (\ref{yq3}) can be written in terms of the Higgs field as
 \begin{eqnarray}
 \frac{\omega(\zeta)}{2}\mathcal D^\nu\zeta\mathcal{D}_\nu\zeta=\frac{\omega_{eff} ({\cal Q})}{2} \left(\partial^ {\nu} {\cal Q}\partial_{\nu}{\cal Q}-\gamma^2\xi\sigma^2A^\nu A_\nu\right.\nonumber\\
 \label{kt1}
 \left.
 -2\gamma^2\sqrt{\xi}\,\sigma {\cal Q} A ^\nu A_\nu-\gamma^2 {\cal Q}^2A^\nu A_\nu\right),
 \end{eqnarray}
 where $\omega_{eff}({\cal Q})=\omega(\sqrt{\xi}\sigma+{\cal Q})$.
 Now, to implement the cosmological principle we make the gauge election: $\theta_{,\mu}=A_{\mu}$ or equivalently $\overline{A}_\mu=0$.  Under this gauge election, the terms in \eqref{kt1} that depend of the electromagnetic field $A_{\mu}$ become null and thus the action \eqref{yq3} becomes
 \begin{equation}\label{newac}
 {\cal S}=\int d^{4}x\sqrt{-h}\left[\frac{R}{16\pi G}+\frac{1}{2}\omega_{eff}({\cal Q})h^{\mu\nu}{\cal Q}_{,\mu}{\cal Q}_{,\nu}-V_{eff}({\cal Q})\right],
 \end{equation}
 where $V_{eff}({\cal Q})=V(\sqrt{\xi}\sigma+{\cal Q})$. In this manner, in order to have a scalar field with a canonical kinetic term we implement the field transformation
 \begin{equation}\label{ft}
 \phi(x^{\sigma})=\int\sqrt{\omega_{eff}({\cal Q})}\,d{\cal Q}.
 \end{equation}
Thus, the action \eqref{newac} in terms of $\phi$ results
\begin{equation}\label{yq4}
{\cal S}=\int d^{4}x\sqrt{-h}\left[\frac{\mathcal{R}}{16\pi G} +\frac{1}{2} h^{\mu\nu} \phi_{,\mu}\phi_{,\nu}-U(\phi)\right],
\end{equation}
where 
\begin{equation}\label{newpot}
U(\phi)=V_{eff}[{\cal Q}(\phi)]=\frac{\lambda}{4}\left[\frac{(\sqrt{\xi}\sigma+{\cal Q}(\phi))^2}{\xi}-\sigma^2\right]^2,
\end{equation}
is the potential associated to the new field $\phi$.
The field equations obtained from the action (\ref{yq4}) read
  \begin{eqnarray}
    G_{\alpha\beta}=-8\pi G[\phi_{,\alpha}\phi_{,\beta}-\frac{1}{2}h_{\alpha\beta}\left(\phi^{,\mu}\phi_{,\mu} -2U(\phi)\right)], \label{Rie6}\\
   \Box\phi+U'(\phi)=0,\label{Rie7}
 \end{eqnarray}
 with $\Box$ denoting the D'Alambertian operator and the prime representing derivative with respect to $\phi$.\\
  
 Thus, in order to consider the Higgs inflationary model developed in \cite{NRI}, we will use the anzats 
\begin{equation}\label{ulca2}
\omega_{eff}(Q)=\frac{1}{\left[1-\beta^2(\sqrt{\xi}\sigma+Q)^4\right]^{5/2}},
\end{equation}
where $\beta$ is a constant parameter with units of $M_{p}^{-2}$. Hence, it follows from \eqref{ft} that
\begin{equation}\label{ulca3}
\phi=\frac{\sqrt{\xi}\sigma+Q}{\left[1-\beta^2(\sqrt{\xi}\sigma + Q)^4\right]^{1/4}}.
\end{equation} 
It can be verified that when $1-\beta^2(\sqrt{\xi}\sigma+Q)^4>0$ 
the ex\-pre\-ssion \eqref{ulca2} is free of pole singularities. Such  condition is fullfilled during inflation.
Therefore the potential \eqref{newpot} acquires the form
\begin{equation}\label{ulca4}
U(\phi)=\frac{\lambda}{4\xi^2}\left(\frac{\phi^4}{1+\beta^2\phi^4}\right).
\end{equation}
It is not difficult to see that the choice of the anzats \eqref{ulca2} allows the effective Higgs potential \eqref{newpot} to exhibit a plateu for large enough field values, making posible a suitable slow-roll inflation. Something similar is used for example in \cite{anz}.
After inflation begins the condition $\beta^2\phi^4\ll 1$ holds, and the potential \eqref{ulca4} can be approximated by
\begin{equation}\label{potre}
U(\phi)\simeq \frac{\lambda}{4\xi^2}\phi^4.
\end{equation}
With the idea in mind to study non-perturbative gauge invariant scalar fluctuations of the metric we will use the non-perturvative formalism introduced in \cite{GIF1}. In this formalism the amplitude of scalar fluctuations is arbitrary. Thus, we consider the perturbed line element 
\begin{equation}\label{npf1}
ds^2=e^{2\psi}dt^2-a^2(t)e^{-2\psi}(dx^2+dy^2+dz^2),
\end{equation}
where $\psi(t,x,y,z)$ is a metric function describing gauge invariant scalar fluctuations of the metric in a non-perturbative manner and $a(t)$ is the cosmic scale factor.\\

Inserting \eqref{npf1} in \eqref{Rie6}, the perturbed field equations read
\begin{eqnarray}
&& e^{-2\psi}\left(3H^2-6H\dot{\psi}+3\dot{\psi}^2\right)+\frac{e^{2\psi}}{a^2}\left[2\nabla\psi-(\nabla\psi)^2\right]\nonumber\\
&& =8\pi G\left[\frac{1}{2}\dot{\phi}^2e^{-2\psi}+\frac{e^{2\psi}}{2a^2}(\nabla\phi)^2+U(\phi)\right],\label{npf2}\\
&& e^{-2\psi}\left(2\ddot{\psi}-5\dot{\psi}^2+8H\dot{\psi}-\frac{2\ddot{a}}{a}-H^2\right)+\frac{e^{2\psi}}{3a^2}(\nabla\psi)^2\nonumber\\
&& =8\pi G \left[\frac{1}{2}e^{-2\psi}\dot{\phi}^2-\frac{e^{2\psi}}{6a^2}(\nabla\phi)^2-U(\phi)\right],\label{npf3}\\
&& \frac{1}{a}\frac{\partial}{\partial x^{i}}\left(\frac{\partial}{\partial t}(a\psi)\right)-\frac{\partial\psi}{\partial t}\frac{\partial\psi}{\partial x^{i}}=4\pi G\dot{\phi}\frac{\partial\phi}{\partial x^{i}},\label{npf4}\\
&& \frac{\partial\psi}{\partial x^{i}}\frac{\partial\psi}{\partial x^{j}}=-4\pi G \frac{\partial\phi}{\partial x^{i}}\frac{\partial\phi}{\partial x^{j}},\label{npf5}
\end{eqnarray}
where the dot is denoting time derivative. 
With the help of \eqref{Rie7} and \eqref{npf1} the equation that determines the dynamics of $\phi$ is given by
\begin{equation}\label{npf6}
\ddot{\phi}+(3H-4\dot{\psi})\dot{\phi}-\frac{e^{4\psi}}{a^2}\nabla\phi+e^{2\psi}U^{\prime}(\phi)=0.
\end{equation}
An algebraic manipulation of \eqref{npf2} and \eqref{npf3} leads to
\begin{eqnarray}
&& e^{-2\psi}\left(4H^2+\frac{2\ddot{a}}{a}-2\ddot{\psi}-14H\dot{\psi}+8\dot{\psi}^2\right)+\frac{e^{2\psi}}{a^2}\left[2\nabla^2\psi\right.\nonumber\\
\label{npf7} 
&&\left.-\frac{4}{3}(\nabla\psi)^2\right] =8\pi G \left[\frac{2}{3a^2}e^{2\psi}(\nabla\phi)^2+2U(\phi)\right],
\end{eqnarray}
which determines the dynamics of the scalar fluctuations of the metric $\psi$.

\section{Gauge invariant scalar fluctuations of small amplitude}

In order to obtain the power spectrum of scalar fluctuations of the metric during inflation it is necessary to consider the impact of quantum amplitudes of $\psi$ on cosmological scales at the end of inflation. Therefore a linear approximation of the equations \eqref{npf4} to \eqref{npf7} will be su\-ffi\-cient to model such small quantum scalar fluctuations $\psi$. Hence, we can use the formula: $e^{\pm n\psi}\simeq 1\pm n\psi$. In this scenario the gauge invariance of $\psi$ can be assu\-red and the weak field limit for the inflaton holds. Thus, it is valid the semiclassical approximation $\phi(t,x^{i})=\phi_{b}(t)+\delta\phi(t,x^{i})$ where $\phi_b(t)=\left<E|\phi|E\right>$ is the background classical field with $|E>$ denotes a physical quantum state determined by the Bunch-Davies vacuum \cite{Bunch-Davies} and $\delta\phi$ describes the quantum fluctuations of the field $\phi$.  \\
In this manner, linearization of the differential line element \eqref{npf1} reads
\begin{equation}\label{lpf1}
ds^2=(1+2\psi)dt^2-a^2(t)(1-2\psi)(dx^2+dy^2+dz^2).
\end{equation}
Analogously, a linearization procedure of the field equations \eqref{npf2} and \eqref{npf3} lead to the classical equations
\begin{eqnarray}
3H^2 &=& 8\pi G\left[\frac{1}{2}\dot{\phi}_b^2+U(\phi_b)\right],\label{lpf2}\\
-2\frac{\ddot{a}}{a}-H^2 &=& 8\pi G\left[\frac{1}{2}\dot{\phi}_b^2-U(\phi_b)\right].\label{lpf3}
\end{eqnarray}
The quantum part obtained from the linearization of \eqref{npf7} is given by
\begin{equation}\label{lpf4}
\ddot{\psi}+7H\dot{\psi}-\frac{1}{a^2}\nabla^2\psi+(6H^2+2\dot{H})\psi=-8\pi G U^{\prime}(\phi_b)\delta\phi.
\end{equation}
With the help of the linearization of \eqref{npf4} and \eqref{npf5} we obtain the relation
\begin{equation}\label{lpf5}
\delta\phi=\frac{1}{4\pi G\dot{\phi}_c}\left(H\psi+\dot{\psi}\right).
\end{equation}
Inserting \eqref{lpf5} in \eqref{lpf4} we arrive to
\begin{eqnarray}
&&\ddot{\psi}+\left(7H`+\frac{2U^{\prime}(\phi_b)}{\dot{\phi}_b}\right)\dot{\psi}-\frac{1}{a^2}\nabla^2\psi+\left(6H^2+2\dot{H}\right.\nonumber\\
\label{lpf6}
&& \left.+\frac{2U^{\prime}(\phi_b)}{\dot{\phi}_b}H\right)\psi=0.
\end{eqnarray}
The linearization of \eqref{npf6} gives the system 
\begin{eqnarray}
&&\ddot{\phi}_b+3H\dot{\phi}_b+U^{\prime}(\phi_b)=0,\label{lpf7}
\\
&& \delta\ddot{\phi}+3H\delta\dot{\phi}-4\dot{\phi}_b\dot{\psi}-\frac{1}{a^2}\nabla\delta\phi+U^{\prime\prime}(\phi_b)\delta\phi\nonumber\\
&& +2U^{\prime}(\phi_b)\psi=0.\label{lpf8}
\end{eqnarray}
Under the slow-roll condition $|\dot{\phi}_b^2/2|\ll |U(\phi_b)|$ it follows from  \eqref{lpf2} and \eqref{lpf7} that
\begin{equation}\label{lpf9}
\dot{\phi_b}=-\frac{M_{p}}{\sqrt{3}}\frac{U^{\prime}(\phi_b)}{\sqrt{U(\phi_b)}},
\end{equation}
where we have taken $M_P=(8\pi G)^{-1/2}$. This equation determines the background inflaton field dynamics.\\

Employing \eqref{ulca4} and \eqref{lpf9} the background field $\phi$ is given by the equation \cite{MMZ}
\begin{eqnarray}
&&t-t_0+\frac{\beta^2}{6\mu}\left(\phi_b^4\sqrt{1+\beta^2\phi_b^4}-\phi_0^4\sqrt{1+\beta^2\phi_0^4}\,\right)+\nonumber\\
&& \frac{2}{3\mu}\left(\sqrt{1+\beta^2\phi_b^4}-\sqrt{1+\beta^2\phi_0^4}\,\right)+\nonumber\\
&& \frac{1}{2\mu}tanh^{-1}\left(\frac{1}{\sqrt{1+\beta^2\phi_b^4}}\right)-\frac{1}{2\mu}tanh^{-1}\left(\frac{1}{\sqrt{1+\beta^2\phi_0^4}}\right)\nonumber\\
&& =0,
\label{lpf10}
\end{eqnarray}
where $\mu=\sqrt{M_p^2\lambda/(3\xi^2)}$ and $\phi_0=\phi(t_0)$, with $t_0$ being the time when inflation begins. Using the potential \eqref{potre} the expression \eqref{lpf10} becomes
\begin{equation}\label{lpfa11}
\phi_b(t)=\phi_e e^{2M_p\sqrt{\frac{\lambda}{3\xi^2}}\,(t_e-t)}.
\end{equation}
where $\phi_e=\phi(t_e)$ with $t_e$ denoting the time when inflation ends. Inserting \eqref{potre} and  \eqref{lpfa11} in \eqref{lpf2} we obtain a scale factor of the form
\begin{equation}\label{lpfa13}
a=a_e\exp \left[\frac{\phi_e^2}{8M_p^2}\left(1-\exp\left(4M_p\sqrt{\frac{\lambda}{3\xi^2}}\,(t_e-t)\right)\right)\right].
\end{equation}
When $t\simeq t_e$ the scale factor \eqref{lpfa13} can be approximated by
\begin{equation}\label{lpfa14}
a(t)\simeq \tilde{a}_e\exp\left(\frac{\phi_e^2}{2M_p}\sqrt{\frac{\lambda}{3\xi^2}}\,t\right),
\end{equation}
where $\tilde{a}_e=a_e\exp\left(-\frac{\phi_e^2}{2M_p}\sqrt{\frac{\lambda}{3\xi^2}}\,t_e\right)$. Thus, the Hubble parameter obtained from \eqref{lpfa13} reads
\begin{equation}\label{lpfa15}
H(t)=\frac{1}{\sqrt{3}M_p}\sqrt{\frac{\lambda}{4\xi^2}}\,\phi_e^2\exp\left(4M_p\sqrt{\frac{\lambda}{3\xi^2}}\,(t_e-t)\right).
\end{equation}
Near the end of inflation, according to \eqref{lpf14}, the Hubble parameter becomes
\begin{equation}\label{lpfa16}
H_e=\left.H\right|_{t\simeq t_e}\simeq \frac{\phi_e^2}{2M_p}\sqrt{\frac{\lambda}{3\xi^2}}.
\end{equation}
On the other hand, Planck data indicate that Higgs inflation requires an energy scale that corresponds to an initial Hubble parameter $H_0\simeq 10^{11}-10^{12}\,GeV$, which is inferred for an average  Higgs mass of the order $M_h\simeq125.7\,GeV$ \cite{Hmass1,Hmass2}. Therefore 
\begin{equation}\label{chp1}
H_0\simeq\frac{\lambda}{2\sqrt{3}}\frac{1}{\beta\xi M_{p}}\simeq 10^{11}-10^{12}\,GeV.
\end{equation}
It is not difficult to verify that for $\lambda=0.13$ and $M_p=1.22\cdot10^{19}\,GeV$ \cite{Hmass2}, the parameter $\xi$ must range in the interval: $\left[3.7528\cdot 10^{-14},3.7528\cdot 10^{-13}\right](\beta M_{p})^{-1}(GeV)^{-1}$. \\

Now, we are in position to quantize the theory. In order to do that we will follow a canonical quantization procedure. Thus, we impose the commutation relation
\begin{equation}\label{lpf11}
\left[\psi(t,\bar{x}),\Pi_{\psi}^0(t,\bar{x}^{\prime})\right]=i\delta^{(3)}(\bar{x}-\bar{x}^{\prime}),
\end{equation}
where $\Pi_{\psi}^0=\partial L/\partial \dot{\psi}$ is the cannonical conjugate momentum to $\psi$ and $L$ denotes the lagrangian given in this case by
\begin{equation}\label{lpf12}
L=\sqrt{-h}\left[\frac{R}{16\pi G}+\frac{1}{2}h^{\mu\nu}\phi_{,\mu}\phi_{,\nu}-U(\phi)\right],
\end{equation}
where $R$ is the Ricci scalar curvature which has the form
\begin{eqnarray}
&& R=\left(6H^2+6\frac{\ddot{a}}{a}-30H\dot{\psi}-9\ddot{\psi}+18\dot{\psi}^2\right)e^{-2\psi}\nonumber\\
\label{lpf13}
&&
+\frac{2}{a^2}(\nabla\psi-(\nabla\psi)^2)e^{2\psi}.
\end{eqnarray}
With the help of equations \eqref{lpf12} y \eqref{lpf13} the relation \eqref{lpf11} becomes
\begin{equation}\label{lpf14}
\left[\psi(t,\bar{x}),\dot{\psi}(t,\bar{x}^{\prime})\right]=i\frac{4\pi G}{9\sqrt{-h}}\,\delta^{(3)}(\bar{x}-\bar{x}^{\prime}).
\end{equation}
To simplify the structure of \eqref{lpf6} we introduce the auxi\-liary field $\zeta$ defined by the formula
\begin{equation}\label{lpf15}
\psi(t,\bar{x})=\exp\left[-\frac{1}{2}\int\left(7H+\alpha\right)dt\right]\zeta(t,\bar{x}),
\end{equation}
where $\alpha=2U^{\prime}(\phi_b)/\dot{\phi}_b$. The equation \eqref{lpf6} in terms of $\zeta$ then reads
\begin{equation}\label{lpf16}
\ddot{\zeta}-\frac{1}{a^2}\nabla^2\zeta-\left[\frac{3}{2}\dot{H}+\frac{1}{2}\dot{\alpha}+\frac{25}{4}H^2+\frac{5}{2}\alpha H+\frac{1}{4}\alpha^2\right]\zeta=0.
\end{equation}
Expanding the field $\zeta(t,\bar{x})$ in Fourier modes we have
\begin{equation}\label{lpf17}
\zeta(t,\bar{x})=\frac{1}{(2\pi)^{3/2}}\int d^3k\left[a_{k}e^{i\bar{k}\cdot\bar{x}}\Theta_k(t)+a_k^{\dagger}e^{-i\bar{k}\cdot\bar{x}}\Theta_{k}^{*}(t)\right],
\end{equation}
where $a_{k}^{\dagger}$ and $a_k$ denote the creation and annihilitation operators obeying the commutation algebra
\begin{eqnarray}\label{lpf18}
\left[a_k,a_{k^{\prime}}^{\dagger}\right]&=&\delta^{(3)}(\bar{k}-\bar{k}^{\prime}),\\
\label{lpf19}
\left[a_k,a_{k^{\prime}}\right]&=& \left[a_{k}^{\dagger},a_{k^{\prime}}^{\dagger}\right]=0.
\end{eqnarray}
The asterisk mark $(*)$ is denoting complex conjugate and the dagger $(\dagger)$ transpose complex conjugate. Thus, it follows from \eqref{lpf16} that the modes $\Theta_k$ obey
\begin{equation}\label{lpf20}
\ddot{\Theta}_k+\left[\frac{k^2}{a^2}-\frac{3}{2}\dot{H}-\frac{1}{2}\dot{\alpha}-\frac{25}{4}H^2-\frac{5}{2}\alpha H-\frac{1}{4}\alpha^2\right]\Theta_k=0.
\end{equation}
With the help of \eqref{lpfa11}, \eqref{lpfa16} and \eqref{ulca4} we obtain
\begin{eqnarray}
&& \alpha=-\frac{6H_e}{\phi_e^3}\left[\frac{\phi_e^3e^{6M_p\sqrt{\frac{\lambda}{3\xi^2}}\,(t_e-t)}}{1+\beta^2\phi_e^4 \,e^{8M_p\sqrt{\frac{\lambda}{3\xi^2}}\,(t_e-t)}}\right.\nonumber\\
&& \left.-\frac{\beta\phi_e^7e^{14M_p\sqrt{\frac{\lambda}{3\xi^2}}\,(t_e-t)}}{\left(1+\beta\phi_e^4e^{8M_p\sqrt{\frac{\lambda}{3\xi^2}}\,(t_e-t)}\right)^2}\right]\,e^{-2M_p\sqrt{\frac{\lambda}{3\xi^2}}\,(t_e-t)}.\nonumber\\
\label{lpf21}
\end{eqnarray}
Thus, at the end of inflation \eqref{lpf21} can be approximated by
\begin{equation}\label{lpf22}
\left.\alpha\right|_{t\simeq t_e}\simeq-\frac{6\gamma_e H_e}{\phi_e^3},
\end{equation}
where 
\begin{equation}\label{lpf24}
\gamma_e=\frac{\phi_e^3}{1+\beta^2\phi_e^4}-\frac{\beta^2\phi_e^7}{\left(1+\beta^2\phi_e^4\right)^2}.
\end{equation}
Using \eqref{lpfa14}, \eqref{lpfa16} and \eqref{lpf22} in \eqref{lpf20} we obtain
\begin{eqnarray}
&&\ddot{\Theta}_k+\left[\frac{k^2}{\tilde{a}_e^2e^{2H_e\,t}}-\frac{25}{4}H_e^2-\frac{5}{2}\alpha_e H_e-\frac{1}{4}\alpha_e^2\right]\Theta_k=0.\nonumber\\
\label{lpf23}
\end{eqnarray}
It follows from \eqref{lpf14}, \eqref{lpf17}, \eqref{lpf18} and \eqref{lpf19} that the nor\-ma\-li\-za\-tion condition is given by
\begin{equation}\label{lpf25}
\Theta_{k}\dot{\Theta}_k^{*}-\Theta_k^{*}\dot{\Theta}_k=i\frac{4\pi G}{9\tilde{a}_e^3}.
\end{equation}
Thus considering a Bunch-Davies vacuum condition the normalized solution of \eqref{lpf23} results to be
\begin{equation}\label{lpf26}
\Theta_{k}(t)=\frac{1}{6M_p}\sqrt{\frac{\pi }{2\tilde{a}_e^3H_e}}\,{\cal H}_{\nu}^{(2)}[z(t)],
\end{equation}
where ${\cal H}_{\nu}^{(2)}[z(t)]$ is the second kind Hankel function, the index $\nu=(-6\gamma_e+5\phi_e^3)/(2\phi_e^3)$ and 
\begin{equation}\label{lpf27}
z(t)=\frac{k}{\tilde{a}_e H_e}\,e^{-H_e\,t}.
\end{equation}
The squared quantum fluctuations of $\psi$ in the IR-sector (on cosmological scales) are given by
\begin{equation}\label{lpf28}
\left<\psi^2\right>=\frac{1}{2\pi^2}e^{-\int{(7H+\alpha)}dt}\int_{0}^{\epsilon k_h}\frac{dk}{k}\,k^3\left.\Theta_{k}\Theta_k^{*}\right|_{IR},
\end{equation}
where $\epsilon=k_{max}^{IR}/k_p\ll 1$ is a dimensionless parameter, $k_{max}^{IR}=k_h(t_r)$ is the wave number related to the Hubble radius at the time when the modes re-enter the horizon $t_r$ and $k_p$ is the Planckian wave number. For a number of e-foldings $N=63$ the parameter $\epsilon$ ranges between $10^{-5}$ and $10^{8}$, which corresponds to a Hubble parameter at the end of inflation of order $H_e=0.5\cdot 10^{-9}M_p$. \\

At the end of inflation, on cosmological scales, we can use  ${\cal H}_{\nu}^{(2)}[z]\simeq (i/\pi)\Gamma(\nu)(z/2)^{-\nu}$. Thus  according to \eqref{lpf28} and \eqref{lpf26} we obtain
\begin{equation}\label{lpf29}
\left<\psi^2\right>=\frac{2^{2\nu-4}}{9\pi^3}\frac{\Gamma^2(\nu)}{M_p^2}\frac{H_e^2e^{((2\nu-7)H_e-\alpha_e)t}}{(\tilde{a}_eH_e)^{3-2\nu}}\int_{0}^{\epsilon k_h}\frac{dk}{k}k^{3-2\nu},
\end{equation}
where $k_h=\tilde{a}_e\sqrt{(25/4)H_e^2+(5/2)\alpha_eH_e+(1/4)\alpha_e^2}$. Hence, the corresponding power spectrum reads
\begin{equation}\label{lpf30}
{\cal P}_s(k)=\frac{2^{2\nu-2}}{9\pi}\frac{\Gamma^2(\nu)}{M_p^2}\left(\frac{H_e}{2\pi}\right)^2e^{((2\nu-7)H_e-\alpha_e)t}\left(\frac{k}{\tilde{a}_e H_e}\right)^{3-2\nu}.
\end{equation}
Notice that for nearly scale invariant: $\nu\simeq 3/2$ it fo\-llows from \eqref{lpf30} that ${\cal P}_s(k)|_{\nu\simeq 3/2}\simeq H_e^2/4\pi^2$. The scale inva\-riance is achieve when $\gamma_e\simeq (1/3)\phi_e^3$. This condition leaves to 
\begin{equation}\label{lpf31}
\frac{1}{1+\beta^2\phi_e^4}-\frac{\beta^2\phi_e^4}{(1+\beta^2\phi_e^4)^2}\simeq\frac{1}{3}.
\end{equation}
Solving this equation we obtain that it is satisfied for the value: $\phi_e=[(\sqrt{3}-1)\beta^2]^{1/4}/\beta$. The spectral index is then $n_s=4-2\nu=(6\gamma_e/\phi_e^3)-1$. Hence it is not difficult to show that
\begin{equation}\label{lpf32}
n_s=6\left[\frac{1}{1+\beta^2\phi_e^4}-\frac{\beta^2\phi_e^4}{(1+\beta^2\phi_e^4)^2}\right]-1.
\end{equation}
The Planck 2018 observational results indicate that the spectral index ranges in the interval $n = 0.968 ± 0.006$ \cite{PRe}.
It follows from \eqref{lpf32} that the inflation field at the end of inflation in terms of $n_s$ is given by the formula
\begin{equation}\label{lpf33}
\phi_e=\frac{1}{\sqrt{\beta}}\left[\frac{6}{\sqrt{6(1+n_s)}}-1\right]^{1/4}
\end{equation}
In this manner, we obtain $\phi_e< M_p$ when the condition  $\beta >\left[(6/\sqrt{6(1+n_s)}\,)-1\right]^{1/2}\,M_p^{-2}$ holds. For example when $n_s=0.9735$ the previous condition reduces to $\beta > 0.8623 M_{p}^{-2}$.\\

On the other hand, the scalar to tensor ratio is given by \cite{MMZ}
\begin{equation}\label{lpf34}
r\simeq \frac{128}{\beta^{2/3}M_p^{4/3}}\frac{1}{(24N)^{5/3}},
\end{equation}
where $N$ is the number of e-foldings at the end of inflation. For $N=63$ we obtain that $r<0.10$, as indicated by Planck observations \cite{PRe}, when $\beta >5.15\cdot 10^{-4}\,M_p^{-2}$. Thus for the aforementioned limit $\beta > 0.8623 M_{p}^{-2}$ the condition $r<0.10$ can be perfectly satisfied. For example,
for $\beta=0.9\,M_p^{-2}$ we obtain $r=6.8\cdot 10^{-4}$.
For this par\-ti\-cu\-lar case we obtain from \eqref{lpf33} that $\phi_e=0.97\,M_p$. For $\beta=16.29053 M_p^{-2}$ we obtain a  scalar to tensor ratio $r=1\cdot 10^{-4}$. This value corresponds to $\phi_e=0.23\, M_p$. Thus the transplanckian problem is avoided in this model.

\section{Final Remarks}

In this letter we have studied gauge invariant fluctuations of the metric in the framework of a recently proposed Higgs inflationary model were the Higgs field has a geometrical origin.
The model has been developed in the theoretical context of a geometrical complex scalar-tensor theory of gravity in which the scalar field form part of the affine structure of the space-time manifold and the gravitational field has a scalar and tensor components. The background geometry was determined via a Palatini variational principle. The description is made from two equivalent frames related by the Weyl transformations in such manner that the Ricci tensor remains unaltered avoiding in this way the unitarity problem \cite{NRI,UP1}. The Higgs scalar field plays the role of the inflaton field and it has its origin from the geometrical Weyl scalar field by means of a particular Weyl transformation. In the model the original Higgs potential is rescaled by the non-canonnical kinetic function $\omega({\cal Q})$ associated to the Weyl-scalar field, physically making  possible to have the enough energy to inflate the universe. We have considered an ansatz for the $\omega({\cal Q})$ function in order to create the enough plateu for the inflationary potential to achieve an energy scale for Higgs inflation corresponding to an initial Hubble parameter $H_0\simeq 10^{11}-10^{12}\, GeV$, which is in agreement with the requirements of PLANCK data for this kind of inflation \cite{Hmass1,Hmass2}.\\

In order to study gauge invariant scalar fluctuations of the metric we started obtaining the dynamical field equations \eqref{npf2}-\eqref{npf5} for the metric function $\psi$ that describes the aforementioned fluctuations in a non-pertubative a\-pproach, as the one described in \cite{GIF1}. As a particular case we have focused in the linear fluctuations and we obtain a nearly scale invariant spectrum at the end of inflation when $\gamma_e\simeq (1/3)\phi_e^3$. We get a scalar index $n_s=0.9735$ and a scalar to tensor ratio $r=6.8\cdot 10^{-4}$ when $\beta=0.9\,M_p^{-2}$. Moreover, for $\beta=16.29053 M_p^{-2}$ we obtain a  scalar to tensor ratio $r=1\cdot 10^{-4}$. This value corresponds to $\phi_e=0.23\, M_p$. Thus, we can say that our model fits well with the Planck 2018 observational data related with the inflationary epoch. In addition, these values for $\phi_e$ indicate that this model is free of the transplanckian problem. Finally, we would like to mention that there are several inflation models with very small value of the tensor to scalar ratio $r$, as for example K$\ddot{a}$hler-moduli and some D-brane, inflationary models. In fact, it may be  expected a next generation of cosmological observational data with a best defined uncertainty for $r$, with upper limits shorter than the have nowadays, $r<0.002$ ($95 \% \,C.L.$) \cite{UREF}

\section*{Acknowledgements}

\noindent  J.E.Madriz-Aguilar, J. Zamarripa and M. Montes  acknowledge CONACYT
M\'exico, Centro Universitario de Ciencias Exactas e Ingenierias and Centro Universitario de los Valles of Universidad de Guadalajara for financial support. A. Bernal thanks Centro Universitario de Ciencias Exactas e Ingenierias for financial support.

\end{document}